\documentclass[10pt,onecolumn,prb]{revtex4-1}
\usepackage{graphicx} % For including figures
\usepackage{bm} % bold math
\usepackage[usenames,dvipsnames]{xcolor}
\usepackage{fullpage}
\usepackage{setspace}
\usepackage{amsmath}
\usepackage{cmbright}
\usepackage[normalem]{ulem}

\usepackage{subfig}
\newcommand{\E}[1]{\ensuremath{{10^{#1}}}}
\renewcommand{\exp}[1]{\ensuremath{{\mathrm{e}^{#1}}}}
\newcommand{\ii}{{\ensuremath{\mathrm{i}}}}
\renewcommand{\vec}[1]{{\ensuremath{\mathbold{#1}}}}
\newcommand{\remark}[1]{}

\begin{document}
\title{Two magnon bound state causes ultrafast thermally induced magnetisation switching}

\author{J.~Barker$^{*}$}
\affiliation{Department of Physics, University of York, York YO10 5DD, U.K.}
\author{U.~Atxitia}
\affiliation{Department of Physics, University of York, York YO10 5DD, U.K.}
\affiliation{Departamento de Fisica de Materiales, Universidad del Pais Vasco, UPV/EHU, 20018 San Sebastian, Spain}
\author{T.A.~Ostler}
\affiliation{Department of Physics, University of York, York YO10 5DD, U.K.}
\author{O.~Hovorka} 
\affiliation{Department of Physics, University of York, York YO10 5DD, U.K.}
\author{O.~Chubykalo-Fesenko}
\affiliation{Instituto de Ciencia de Materiales de Madrid, CSIC, Cantoblanco, 28049 Madrid, Spain.}
\author{R.W.~Chantrell}
\affiliation{Department of Physics, University of York, York YO10 5DD, U.K.}

\begin{abstract}
    There has been much interest recently in the discovery of thermally induced
    magnetisation switching, where a ferrimagnetic system can be switched
    deterministically without and applied magnetic field. Experimental results
    suggest that the reversal occurs due to intrinsic material properties, but
    so far the microscopic mechanism responsible for reversal has not been
    identified. Using computational and analytic methods we show that the
    switching is caused by the excitation of two magnon bound states, the
    properties of which are dependent on material factors.  This discovery
    allows us to accurately predict the switching behaviour and the identification
    of this mechanism will allow new classes of materials to be identified or
    designed to use this switching in memory devices in the THz regime.
\end{abstract}

\maketitle

%\section{Introduction}

Thermally induced ultrafast magnetisation switching (TIMS) is a recent
discovery in which an applied sub-picosecond heat pulse causes the magnetic
state of a system to switch without any symmetry breaking magnetic field
\cite{Ostler:2012hx}.  The lack of any external directional stimulus has long
been intriguing and implies the mechanism to be fundamentally intrinsic to the
specific class of materials in which it is found.  The full microscopic
understanding is still
lacking\cite{Mentink:2012ga,Atxitia:2012vy,Schellekens:2013wg}, and it is not
known why switching has only been observed in the ferrimagnets
GdFeCo\cite{Ostler:2012hx} and TbCo\cite{Alebrand:2012gf}.  Here we study
the phenomenon in GdFeCo using atomistic spin dynamics, supported by
analytical calculations based on the linear spin wave theory. We reveal that
TIMS results from the strong excitation of two magnon bound states,
a hybridisation of ferro- and antiferro-magnetic dynamical modes, relating to
specific material dependent conditions. We give detailed quantification of the
phenomenon, and thus our study opens pathways for search and design of new
classes of materials exhibiting TIMS.

%\section{Main}

The manipulation of magnetism by using ultrafast external stimuli ($<100$
ps)\cite{Stohr:2006wk}, such as shaped magnetic field pulses
\cite{Tudosa:2004jc}, acousto-magneto-plasmonics \cite{Temnov:2012ve} and
femtosecond laser pulses\cite{Bigot:2009bu,Kirilyuk:2010vd}, is fundamental to
future digital data storage technologies\cite{Weller:1999ia}.  The most
promising is the discovery of sub-picosecond magnetisation reversal by TIMS,
occurring after the application of a femtosecond laser pulse alone in
antiferromagnetically coupled systems\cite{Stanciu:2007vh}.  It opens new
avenues for technological developments including proposals for ultrafast,
all-optically switched magnetic recording media which will allow a considerable
simplification in the design of write transducers and achieve significant
energy savings \cite{Challener:2009do,Savoini:2012ik}.  However, despite
extensive experimental
\cite{deJong:2012kh,Alebrand:2012il,Alebrand:2012gf,Savoini:2012ik,Khorsand:2012co,Khorsand:2013cy,Graves:2013ub}
and theoretical research
\cite{Schlickeiser:2012ew,Mentink:2012ga,Atxitia:2012vy,Schellekens:2013wg}
aimed at revealing the microscopic mechanisms to allow the control of TIMS and
the identification of new candidate materials, a satisfactory understanding
still does not exist.  Shedding new light on the issue is the main subject of
this report.

\section*{Results}
\begin{figure}[!h]
\begin{center}
\subfloat{\includegraphics[width=0.45\textwidth]{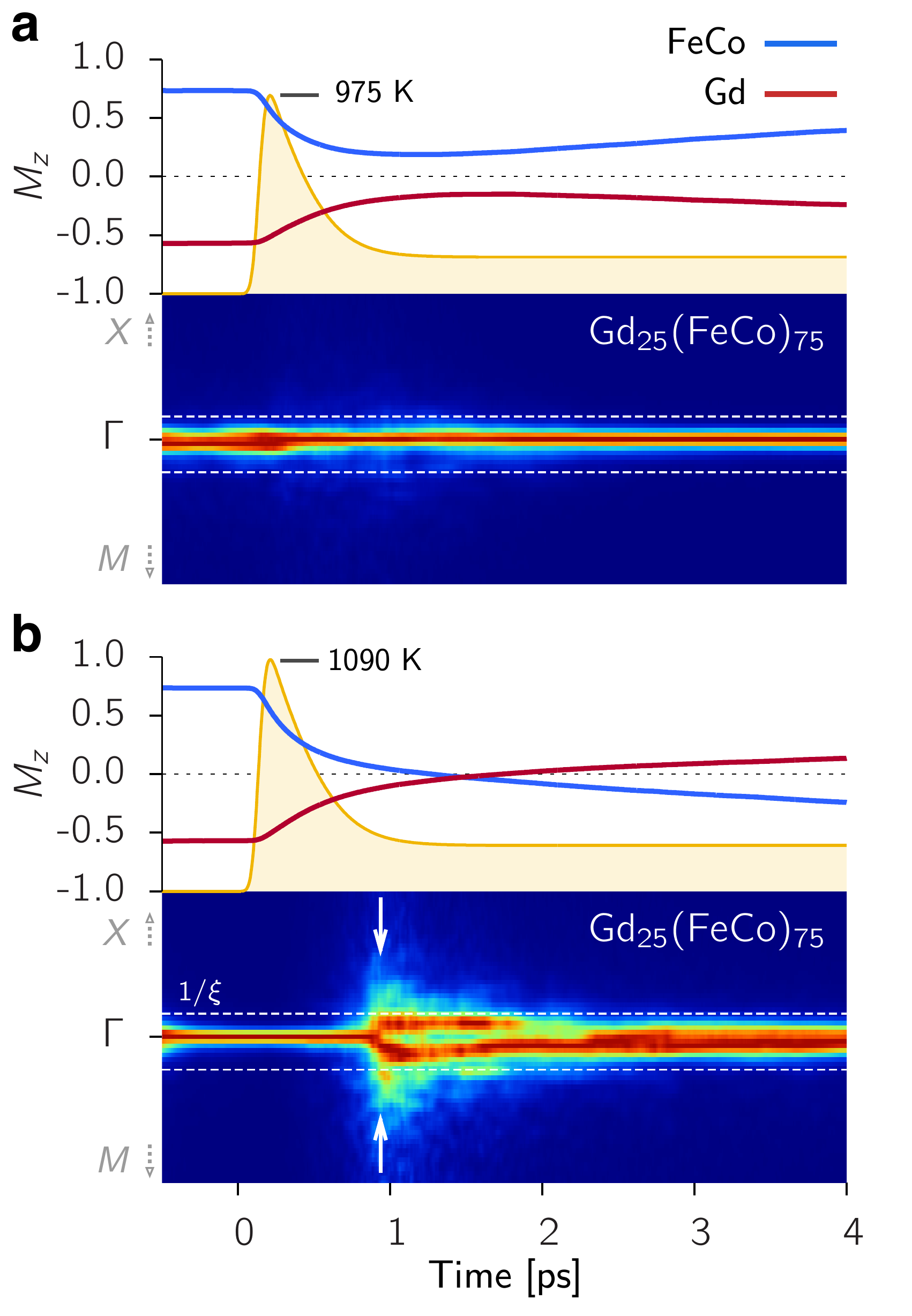}}\\
\end{center}
\caption{
\textbf{Intermediate structure factors with laser excitation.} 
Application of a laser pulse to the Gd$_{25}$(FeCo)$_{75}$ amorphous lattice model. The blue and red lines are the $M_z$
of the FeCo and Gd, normalised to the total magnetisation of each sublattice respectively. The yellow curve shows the thermal energy applied to the system. In the low panels the colour intensity represents the amplitude of magnons at the given $k$-vector in comparison to other vectors at the same instance in time. 
\textbf{(a)} The laser heating causes a reduction in magnetisation of the two sublattices but the distribution of power in the magnons does not change significantly from the equilibrium distribution.
\textbf{(b)} A higher laser fluence causes switching. During the reversal period, magnons on a specific length scale are excited corresponding to the angular momentum transfer channel between AF and FM modes. After reversal the ISF returns to the equilibrium distribution.
}
\label{fig:ssf}
\end{figure}

To understand the microscopic magnetisation dynamics which lead to TIMS in
GdFeCo alloys we first use large-scale atomistic spin dynamics to study time
evolution of the spatial Fourier transform of the spin-spin correlation
function (the intermediate structure factor - ISF) (see Supplementary Section S2) after the
application of a femtosecond laser pulse to observe the distribution of magnons
in the Brillouin zone.  Figure \ref{fig:ssf}a corresponds to a low laser
fluence situation where TIMS is not observed.  The ISF shows that the absorbed
laser energy ($k_BT(t)$) is uniformly distributed within the low wave-vector
$k$ modes after the initial heating, leading to a decrease in the magnetisation
of both sublattices (upper panel Fig.~\ref{fig:ssf}a). After the pulse, the
non-equilibrium magnon distribution moves back towards its equilibrium
resulting in the gradual recovery in the magnetisation of the sublattices. For
a larger laser fluence (Fig.~\ref{fig:ssf}b) the initial heating  leads to
a more pronounced reduction in magnetisation and excitation of a broader
$k$-range.  While cooling, instead of a relaxation of magnons, one observes an
almost instantaneous excitation of magnons  within a  well defined range in
$k$-space. This behaviour is a consequence of magnon-mediated angular momentum
transfer between ferromagnetic (FM) and antiferromagnetic (AF) modes through
well-defined channels.  This initially leads to the so-called transient
ferromagnetic-like state\cite{Ostler:2012hx} which precedes TIMS.  

To identify the nature of magnons defining the angular momentum channels, we
calculate the dynamic structure factor (DSF),
$\mathcal{S}\left(\vec{k},\omega \right)$, to obtain the magnon spectrum
(experimentally obtainable via Brillouin experiments). To interpret the ISF in relation to the equilibrium spectrum, one can assume 
the laser excitation is so fast that only the magnonic
population is altered, rather than the spectrum itself. To give a clear contrast of the
relative contribution  to the spin fluctuations of each magnon branch in the
spectrum we perform the normalisation
$|\mathcal{S}(\vec{k},\omega)|^2/\mathrm{max}(|\mathcal{S}(\vec{k},\omega)|^2)$
\cite{Bergman:2010wb}.

\begin{figure*}[!h]
    \begin{center}
        \subfloat{\includegraphics[width=\textwidth]{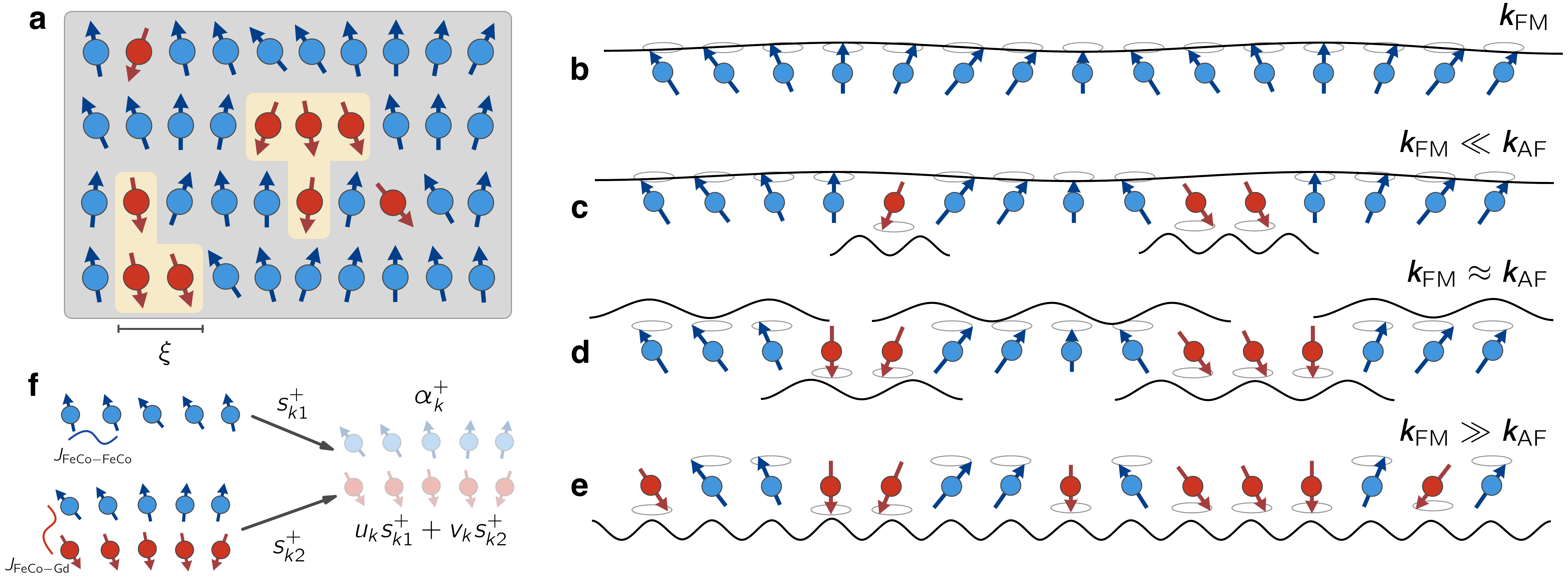}}\\
    \end{center}
    \caption{
        \textbf{Effective magnons in amorphous GdFeCo.} 
        \textbf{(a)} The lattice is made up of FeCo (blue) and Gd (red) spins.
        The random distribution of Gd within the lattice forms clusters of
        connected Gd regions. These have a typical length scale $\xi$ for
        a given Gd concentration.  \textbf{(b-e)} Relevant magnons on different
        wave-vectors $k$, depend on the Gd content. \textbf{(b)} For a pure FM
        only FM magnons exist on all wave-vectors $k$. \textbf{(c)} At low Gd
        concentrations FeCo spin fluctuations are dominated by low $k$ FM
        magnons, whereas the few Gd spins fluctuate due to the short range AF
        coupling to the FeCo lattice represented by the dominance of AF magnons
        at large $k$. \textbf{(d)} At intermediate Gd concentrations ($\sim$
        20-30 \%) due to the increase of Gd spins in the system, the
        AF correlation length characterised by the wave-vector
        $k_{\mathrm{AF}}$ decreases allowing
        the existence of AF-FM two magnon bound state spin fluctuations.
        \textbf{(e)} AF magnons dominate across all length scales.
        \textbf{(f)} Using linear spin wave theory and the virtual crystal
        approximation we can represent the system in terms of $\alpha$ and
        $\beta$ effective magnons which hybridizes the elementary FM and AF
        modes (see Supplementary Section S4 for details).
    }
\label{fig:intro}
\end{figure*}

The spectrum in ferrimagnetic GdFeCo alloys, (Fig.~\ref{fig:intro}a),
contains two kinds of magnons: 1) FM magnons whose low-energy
reads $\omega(k)\sim k^2$\cite{} (see Fig.~\ref{fig:intro}b)  and 2) the
AF magnons ($\omega \sim k$) which relate mainly  to
FeCo-Gd spin-fluctuations \cite{Haug:1972vm,Mekonnen:2011dp}. The magnon
spectrum evolves in a characteristic manner with increasing Gd concentration as
shown schematically in figures.~\ref{fig:intro}c-e. At low density, where the Gd
can be considered an impurity in the FeCo lattice, the spectrum is dominated
mainly by the FeCo spin-fluctuations (Fig.~\ref{fig:intro}c), while at high densities of Gd
the AF mode dominates (Fig.~\ref{fig:intro}e). In the intermediate density
range (Fig.~\ref{fig:intro}d) both FM and AF magnons coexist and, as we will
show, it is this region which is central to the origin of TIMS.  In this work
we use the virtual crystal approximation (VCA) to make the disordered
lattice tractable within linear spin wave theory (LSWT) and calculate
the energy and the contribution of the FM and AF magnons to the spin
fluctuations of each sublattice (Fig.~\ref{fig:intro}f).

\begin{figure*}[!h]
\begin{center}
\includegraphics[width=\textwidth]{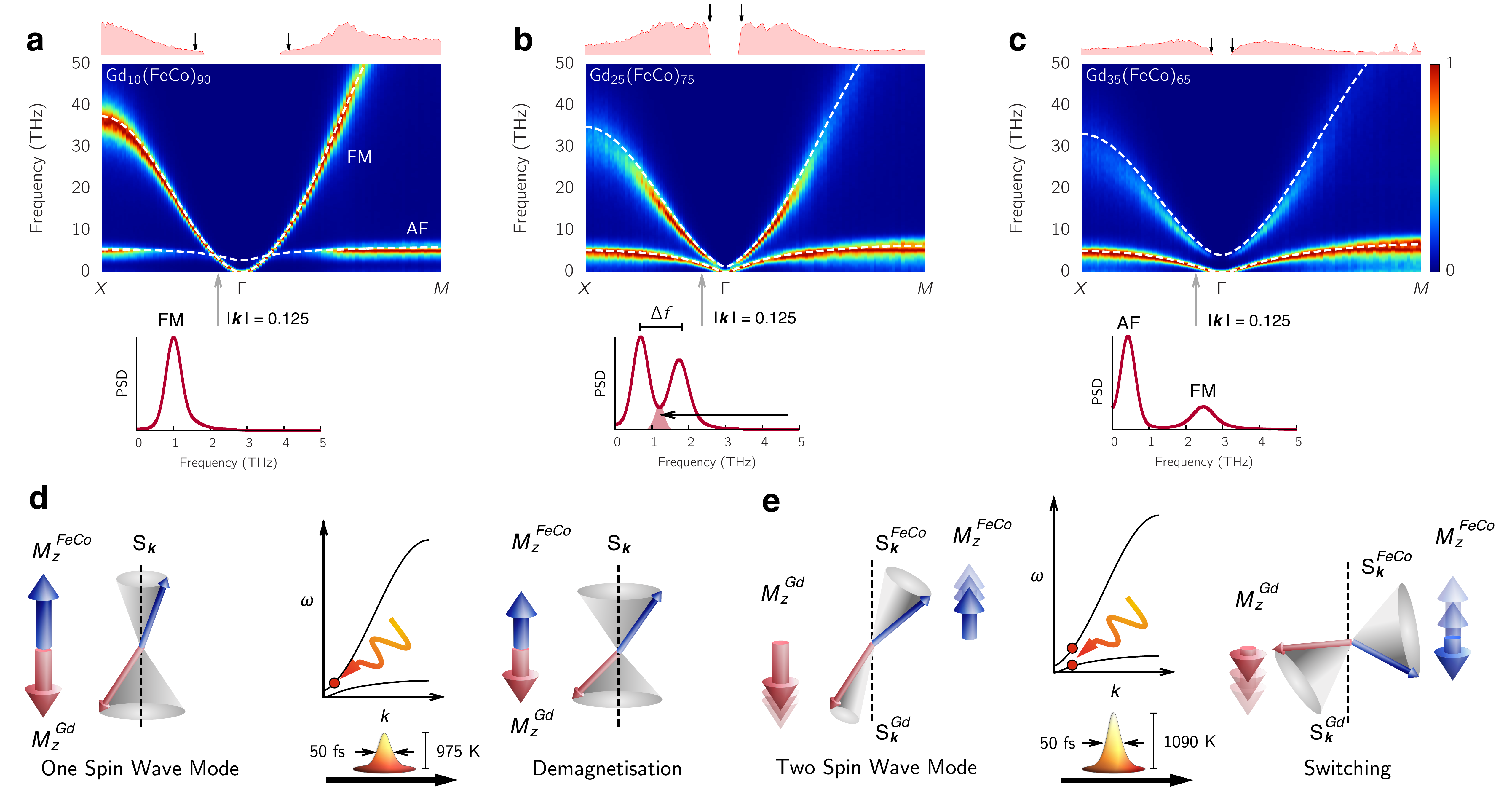}
\end{center}
\caption{
    \textbf{Magnon band structures and explanation of one and two magnon states.} 
    \textbf{(a-c)} In each DSF the colour indicates the relative amplitude of
    magnons modes (the power specturm density).  The analytic dispersion from the LSWT is
    overlaid on each DSF in dashed-white, showing a good agreement with our
    calculated band structure.
    The box above each panel gives the amplitude of the two magnon state in red. This is maximised when both bands have the same amplitude and is zero where only one band contains any
    amplitude. The mean Gd cluster correlation length $\xi$ as calculated from
    percolation theory is denoted by the black arrows. 
    Below each DSF is a cross section of the power spectrum density (PSD) at the
    low $k$ vector $|k|=0.125$.  
    \textbf{(a)} Low Gd concentration has distinct FM and AF magnons.The AF band is restricted to the edge of the Brillouin zone (colloring scheme) as there are relatively few, localised FeCo-Gd interactions. The system
    behaves as a FM due to the dominance of this band. 
    \textbf{(b)}  For Gd$\sim$20-30\%.
    there is region near the centre of the Brillouin zone with the two magnon state
    and a small frequency gap ($\Delta f$) between the two bands. The shaded region
    in the PSD is where non-linear interactions allow the efficient transfer of
    angular momentum between sublattices. Strong excitation of these magnons causes
    TIMS.  
    \textbf{(c)} High Gd concentration reduces the two magnon state and the
    large frequency gap stops the flow of angular momentum between the
    FM and AF modes.
    \textbf{(d)} Excitation of one magnon modes causes a reduction in the magnetisation.
    \textbf{(e)} Two magnon modes cause localised oscillations in the magnetisation. Strong excitation of these states causes a transfer of angular momentum between FM and AF modes leading to the transient ferromagnetic state and switching.
}
\label{fig:dsf_conc}
\end{figure*}

At low Gd concentration, where TIMS is not observed in our simulations or in
experiments \cite{Khorsand:2013cy}, the distinction between FM and AF magnons
is well manifested (Fig~\ref{fig:dsf_conc}a). FM magnons are dominant across
the Brillouin zone, and it is only at the
edge of the Brillouin zone where the few localised FeCo-Gd interactions play
a role.  This indicates the suppression of AF excitations on longer length
scales (small $k$) within the lattice and thus the interaction-induced AF correlation
length is range-limited.  Therefore, in the low-energy regime GdFeCo
essentially behaves as a ferromagnet slightly perturbed by Gd impurities. 
Spin fluctuations are mixed only at very short interaction length scales (large $k$ values) which are very high in energy, so the laser heating excites only FM modes leading to a reduction in magnetisation (Fig.~\ref{fig:dsf_conc}d).

As the Gd concentration increases, so does the FeCo-Gd AF exchange interaction
correlation length(Fig.~\ref{fig:intro}d).  Consequently in the FeCo
lattice the relative FM magnon contribution to spin fluctuations loses
amplitude at large length scales in favour of the AF modes
(Fig.~\ref{fig:dsf_conc}b), gradually diminishing the ferromagnetic character
of such spin fluctuations to a FM-AF magnon mixing, the two magnon bound state.  For 20-30\% Gd, there is
a strong interplay between the two bands and a region
develops close to the centre of the Brillouin zone where the relative amplitude
of both magnon branches  is similar, leading to localised oscillations in the
magnetisation vector that can be excited by the laser energy as shown in
figure~\ref{fig:dsf_conc}e.  This is a key factor to allow angular momentum
transfer between the modes which scales with the intersecting area 
of the two modes power spectrum (see PSD cross section below Fig.~\ref{fig:dsf_conc}b).
This area is maximised when gap between the bands, $\Delta f$ is minimised. 
For TIMS to occur the flow of angular momentum from FM to AF
modes must be enhanced which is satisfied by increasing laser fluence because the
number of magnons transferring angular momentum is increased.

For even larger Gd content ($>35\%$) FeCo-Gd interactions play the dominant
role in the lattice (Fig.~\ref{fig:intro}e). The system takes on the character
of an antiferromagnet with little contribution from the FM band
(Fig.~\ref{fig:dsf_conc}c). The large frequency gap $\Delta f$ means that there
is negligible interaction between two magnon modes, reducing angular momentum
transfer, so applying laser energy causes the system to demagnetise via one magnon
excitations as in figure~\ref{fig:dsf_conc}d.

\begin{figure}[!h]
\begin{center}
\subfloat{\includegraphics[scale=0.9]{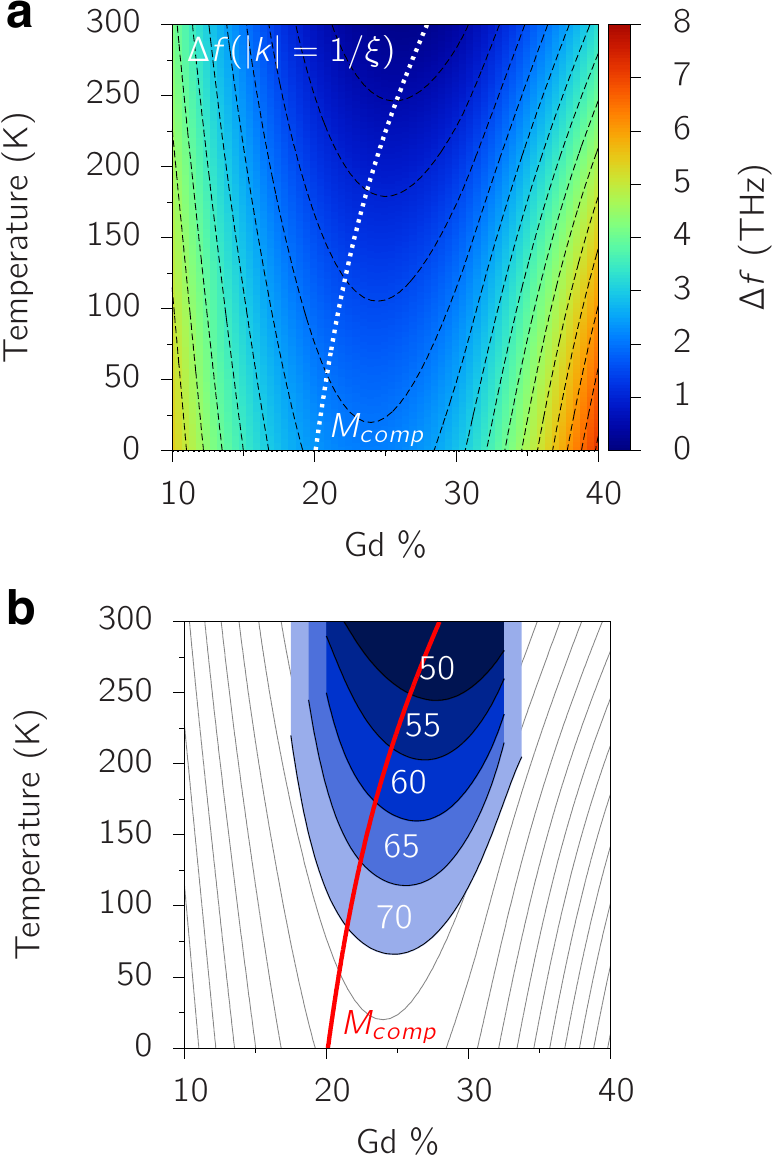}}
\end{center}
\caption{
\textbf{TIMS reversal windows.}  
\textbf{(a)} The band frequency difference $\Delta f$ at the cluster correlation length $\xi$, calculated from LSWT with VCA and percolation theory. The spacial localisation of
the Gd clusters means the minimum does not lie on $M_{comp}$. Smaller $\Delta f$ means that the two magnon modes can more efficiently transfer angular momentum and magnetisation between sublattices when sufficiently excited.
\textbf{(b)} TIMS switching windows found from atomistic spin dynamics for
different laser fluences. Each enclosed area is the parameter set where
switching occurs for the labelled laser fluence. The switching windows closely
match the energy contours from diagram \textbf{(a)} (shown in grey).}
\label{fig:maps}
\end{figure}

The minimum laser energy required to initiate switching is essential in the
interpretation of experiments\cite{Vahaplar:2009fp,Khorsand:2012co}. Laser
induced magnetisation switching was first observed to depend on the chirality
of the laser pulse \cite{Stanciu:2007fy} but was later shown using linear
light\cite{Ostler:2012hx}. The threshold energy is
helicity dependent because of magnetic circular dichroism
\cite{Khorsand:2012co}.  We propose the following criterion for TIMS. First,
the frequency gap $\Delta f(k)$ between both magnon branches should be
minimised, to maximise the angular momentum transfer through nonlinear
interactions (Fig.~\ref{fig:dsf_conc}b). Second, the laser energy must be
sufficient to strongly excite the $k$-region corresponding to the two magnon
bound states. To define this region
we use percolation theory to quantify the statistical properties of
clusters of Gd on the lattice (Fig.~\ref{fig:intro}a) and find the typical
correlation length $\xi$ of clusters (see Methods section). The significance is
that $\xi$ directly relates to the length scale of the AF FeCo-Gd interactions.
We plot $1/\xi$ as dashed white
lines in Fig.~\ref{fig:ssf} where it matches the excited region during switching and as black arrows in the top panels in Fig~\ref{fig:dsf_conc}a-c where it matches the extent of the two magnon state in the DSF, shown in red
above each panel.  From our analytic framework incorporating the LSWT with
the VCA and using $\xi$ calculated via percolation theory, we can
calculate $\Delta f (1/\xi)$, the frequency difference of the low energy two magnon states, for a range of equilibrium
temperatures, $T=0-300$ K and Gd concentrations, $\%=10-40$
(Fig.~\ref{fig:maps}a). To test our premiss that the threshold laser energy
to induce TIMS scales with $\Delta f (1/\xi)$ we perform many computational
simulations as in Fig.~\ref{fig:ssf} to find the parameter regions where TIMS
occurs. Fig.~\ref{fig:maps}b) shows the switching regions for different laser fluences (as labelled). The criterion $\Delta f (1/\xi)$
is smallest around Gd concentrations of $25\%$, but does not exactly
follow the magnetisation compensation point, $M_{\mathrm{comp}}$, where the
magnetisation of each sublattice is equal. This deviation directly relates to
the Gd clustering which limits the range of the two magnon states.  In larger
samples the FeCo clustering around  Gd rich regions  can produce the inverse
effect, namely, showing a transfer of angular momentum to FeCo clusters with
the consequence of Gd region reversing first \cite{Graves:2013ub}.

\section*{Discussion}
Our study has identified the nature of TIMS as the magnon-mediated
angular momentum transfer between ferromagnetic subsystems with
antiferromagnetic coupling between them.  Our quantitative analysis opens the
door for design of magnetic heterostructures for more energy-efficient
all-optical storage devices \cite{Stanciu:2007fy}, and an enhancement of the
information processing rates into the elusive THz regime \cite{Kimel:2007fe}.
The angular momentum transfer channels identified in this work as being
essential for the occurrence of TIMS, can be directly accessed by THz radiation
\cite{Kampfrath:2010kl,Wienholdt:2012cm}. Operation in the THz range leads to
a range of benefits as it allows substantially reducing the heat generation
that leads to material fatigue and device performance degradation.
Additionally, due to the problem with sourcing the rare-earth materials, the
large-scale technological impact  relies on the discoveries of new
cost-friendly TIMS-exhibiting materials. The relatively small parameter space
necessary for existence of TIMS in natural materials such as the GdFeCo alloys
can be broadened via engineering of heterostructures, for instance,
superlattices made of ferromagnetic layers with strong AF coupling, and by
improving the inter-lattice magnon-exchange efficiency.

\section*{Methods}
\subsection*{Lattice Impurity Model}
We model the GdFeCo on a simple cubic lattice with Gd moments placed on sites
with a uniform random probability. Other sites are considered as an effective
FeCo combined moment. The simulations are performed with a Heisenberg like
Hamiltonian 
\begin{equation}
    \mathcal{H} = 
    -\sum_{\langle ij \rangle} J_{ij}^{\mathrm(FeCo-FeCo)} \vec{S}_{i}\cdot\vec{S}_{j}
    -\sum_{\langle ij \rangle} J_{ij}^{\mathrm(Gd-Gd)} \vec{S}_{i}\cdot\vec{S}_{j}
    -\sum_{\langle ij \rangle} J_{ij}^{\mathrm(FeCo-Gd)}\vec{S}_{i}\cdot\vec{S}_{j} 
    - \sum_{i}D_{z}S_{z,i}^2
\label{eq:HamiltonianMethods}
\end{equation}
The laser heating is modelled using a two-temperature model representing the
coupled phonon and electron heat baths. The spin degrees of freedom are coupled
to the electronic temperature.  Calculation of the dynamic structure factor was
by means of a three dimensional spacial discrete Fourier transform (with
periodic boundaries) and temporal discrete Fourier transform where a Hamming
window is applied. We use a simple cubic lattice of size
$128\times128\times128$ and integrate the Landau-Lifshitz-Gilbert Langevin
equation for over 800ps of simulated time, giving a frequency resolution of
2.5GHz. The resulting power spectra are then convoluted along constant
  $k$-vector with a Gaussian kernel of width $\sim 0.95$THz and normalised
  so the largest peak is unity (an example is given in Supplementary Fig.~1).

\subsection*{Linear Spin Wave Theory}
We first use the virtual crystal approximation  to make the disordered lattice
Hamiltonian in Eq. \eqref{eq:HamiltonianMethods} translationally symmetric
with respect to spin variables.  The spin dynamics is described by the linear
Landau-Lifshitz equation of motion, $\mathrm{d}
\mathbf{s}_i/\mathrm{d}t=\gamma[\mathbf{s}_i\times
\mathbf{H}_{\mathrm{eff},i}]$, where $\mathbf{H}_{\mathrm{eff},i}=-\partial
\mathcal{H}/\partial \mathbf{s}_i$. The resulting  equations are then
transformed in terms of spin raising and lowering  operators
$s^{\pm}_{i,\mathrm{FeCo}}=s_{i,\mathrm{FeCo}}^{x}\pm s_{i,\mathrm{FeCo}}^{y}$ and $s^{\pm}_{i,\mathrm{Gd}}=s_{i,\mathrm{Gd}}^{x}\pm s_{i,\mathrm{Gd}}^{y}$
which describes the spin fluctuations around equilibrium.  The resulting system
of two coupled equations is then Fourier transform to describe the spin
fluctuations in the reciprocal space and diagonalised by  Bogoliubov-like
transformation
$\mathbf{s}_{\mathbf{k},\mathrm{FeCo}}=u^+_{\mathbf{k}}\alpha_{\mathbf{k}}+u^{-}_{\mathbf{k}}\beta^{\dag}_{\mathbf{k}}$.
$\mathbf{s}_{\mathbf{k},\mathrm{Gd}}=u^-_{\mathbf{k}}\alpha^{\dag}_{\mathbf{k}}+u^{+}_{\mathbf{k}}\beta_{\mathbf{k}}$,
where $\alpha_{\mathbf{k}}$ and $\beta_{\mathbf{k}}$  are the eigenstates
(magnons) of the system with  frequency $\omega_{\alpha}(\mathbf{k})$ and
$\omega_{\beta}(\mathbf{k})$ respectively.  More detail is given Supplementary Section S4.

\subsection*{Percolation Theory}
Percolation theory provides a general mathematical toolbox for quantifying
statistical properties of connected geometrical regions of size $s$ which will
here refer to $s$ adjacent Gd atom sites. After identifying such Gd clusters
within the lattice using the efficient Hoshen-Kopelman
algorithm\cite{Hoshen:1997ga}, discounting small clusters ($s<4$) and
percolating clusters spanning the computational cell, we calculate the radius
of gyration $R_{st}$ of each cluster remaining within the distribution and
obtain the correlation length as:
\begin{equation}
\xi^{2} = \frac{2\sum_{s}s^{2}\sum_{t=1}^{n_{s}}R_{st}^2}{\sum_{s}s^{2}n_{s}}	
\end{equation}
The finite size effects are included via the finite size scaling formula for
the correlation length 
\begin{equation}
\tilde{\xi} = A |p-p_{c}|^{-1/\nu}	
\end{equation}
where $p_c$ is the percolation threshold for bulk lattice and $\nu$ is
correlation length universal critical exponents. The values $p_c = 0.3116004$
and $\nu=0.875$ for site percolation on a simple cubic lattice and the
non-universal constant $A=0.776187$ obtained by fitting Eq. 3 to the cluster
data evaluated by statistical counts through the lattice. Thus Eq. 3 allows
relating the Gd concentration to the associated typical geometrical size of Gd
clusters, and correlates well with the predictions of the LSWT discussed in the
text.

\newpage

\section*{References}
%\bibliography{/Users/joe/Documents/biblio}

\newpage 

\section*{Supplementary Information}
\subsection*{S1 - Atomistic Spin Model}

The atomistic modelling used to obtain the dynamic structure follows standard
techniques in this area. We include a description of the model here for
completeness. We use the Landau-Lifshitz-Gilbert equation
\begin{equation}\label{eq:landau-lifshitz-gilbert} 
    \frac{\partial \vec{S}_{i}}{\partial t} =
    -\frac{\gamma_{i}}{(1+\alpha_{i}^2)\mu_{i}}\left( \vec{S}_{i} \times \vec{H}_{i} + \alpha_{i}
\vec{S}_{i}\times\vec{S}_{i}\times\vec{H}_{i}\right) 
\end{equation}
where $\gamma_{i}$ is the gyromagnetic ratio, $\alpha_{i}$ is the Gilbert
damping, $\mu_{i}$ is the magnetic moment and $\vec{H}_{i}$ is the effective
field on a spin $\vec{S}_{i}$. We can include temperature by writing the LLG as
a Langevin equation, where the effective field contains a stochastic process
$\vec{\xi}_{i}$
\begin{equation} 
    \vec{H}_{i} = \vec{\xi}_{i} -\frac{\partial \mathcal{H}}{\partial \vec{S}_{i}} 
\end{equation}
the moments of which are defined as
\begin{align}
    \begin{split}\label{eq:landau-lifshitz-langevin-moments}
        \langle \xi_{i}(t) \rangle &= 0\\
        \langle \xi_{i,a}(t), \xi_{j,b}(t^{\prime}) \rangle &= 2 k_{B} T
        \alpha_{i}\delta(|t-t^{\prime}|)\delta_{ij}\delta_{ab}
    \end{split}
\end{align}
where $a$ and $b$ are Cartesian components.
The equation of motion is integrated with the Heun scheme using a time step of
$dt$=0.1~fs to ensure numerical stability. The material parameters we use in the model are given in table~\ref{tab:parameters}.

\begin{table}[htdp]
    \begin{center}
        \begin{tabular}{l c r l r}
            \hline
            FeCo-FeCo Exchange Energy & $J_{ij}$ & 6.920 & J$\times\E{-21}$ & \\
            FeCo-Gd Exchange Energy & $J_{ij}$ & -2.410 & J$\times\E{-21}$ & \\
            Gd-Gd Exchange Energy & $J_{ij}$ & 2.778 & J$\times\E{-21}$ & \\
            \hline
            FeCo Anisotropy Energy& $d_{z}$ & 8.072 & J$\times\E{-24}$ & \\
            FeCo Moment & $\mu_{\mathrm{s}}$ & 1.92 & $\mu_{\mathrm{B}}$ & \\
            FeCo Damping & $\alpha$ & 0.02 & & \\
            FeCo Gyromagnetic Ratio & $\gamma$ & 1.00 & $\gamma_{e}$& \\
            \hline
            Gd Anisotropy Energy& $d_{z}$ & 8.072 & J$\times\E{-24}$ & \\
            Gd Moment & $\mu_{\mathrm{s}}$ & 7.63 & $\mu_{\mathrm{B}}$ & \\
            Gd Damping & $\alpha$ & 0.02 & & \\
            Gd Gyromagnetic Ratio & $\gamma$ & 1.00 & $\gamma_{e}$& \\
            \hline
        \end{tabular}
    \end{center}
    \caption{Atomistic material parameters for GdFeCo in the LLG equation.}
    \label{tab:parameters}
\end{table}

The amorphous nature of GdFeCo is modelled by using a simple cubic lattice
model but with random placements of Gd moments within the lattice to the
desired concentration. Using a very large lattice ($128\times128\times128$)
allows use to finely control the concentration and also gives a good ensemble
of clusters.

The thermal effect of the laser is included by use of the two temperature 
model\cite{Chen:2006bo} where the spin system is coupled to the electron temperature.

\subsection*{S2 - Structure Factors}
The intermediate structure factor (ISF) is calculated from
\begin{equation}
    \mathcal{S}(\vec{k},t) = \frac{1}{N}\sum_{\vec{r},\vec{r}^{\prime}}\exp{\ii \vec{k}\cdot(\vec{r}-\vec{r}^{\prime})}C\left(\vec{r}-\vec{r}^{\prime},t \right)
\end{equation}
where N is the number of spins and the spin-spin correlation function,
$C(\vec{r}-\vec{r}^{\prime},t) =  S_{+}(\vec{r},t)S_{-}(\vec{r}^{\prime},t)$.

The dynamic structure factor (DSF) is calculated from
\begin{equation}\label{eq:dsf}
	\mathcal{S}\left(\vec{k},\omega \right)=\frac{1}{N}
	\sum_{\vec{r},\vec{r}^{\prime}}
	\exp{\ii\vec{k}\cdot(\vec{r}-\vec{r}^{\prime})}
	\int\exp{\ii\omega t} C\left(\vec{r}-\vec{r}^{\prime},t\right)\mathrm{d}t
\end{equation}
where $N$ is the number of spins, $C(\vec{r}-\vec{r}^{\prime},t)$ is the spin-spin correlation function
\begin{equation}
	C(\vec{r}-\vec{r}^{\prime},t) = \langle S_{+}(\vec{r},t)S_{-}(\vec{r}^{\prime},0) \rangle
\end{equation}
$S_{+}$, $S_{-}$ are the spin raising and lowing operators and
$\langle\cdots\rangle$ denotes a thermodynamic average.
Numerically the time integral is performed as a discrete Fourier transform in a Hamming window.

\begin{figure}[htbc!]
\centering
\subfloat[Raw PSD data]
	{\label{fig:convolution_input_example}
	\includegraphics[]{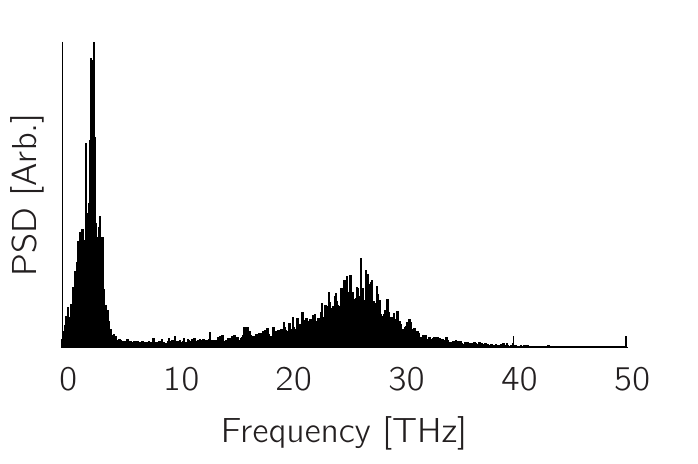}}
\subfloat[PSD data after Gaussian convolution]
{\label{fig:convolution_output_example}
	\includegraphics[]{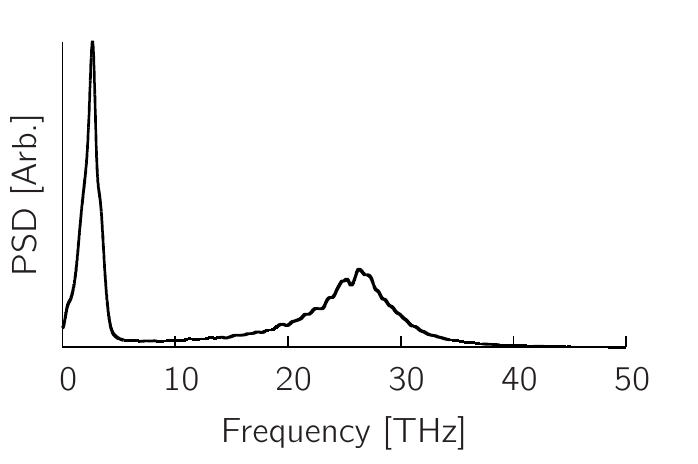}}
    \caption{The power spectral density along one $k$-vector is filtered using 
    Gaussian convolution with a width of $\sigma=0.95$~THz.}
  \label{fig:convolution_example}
\end{figure}

To reduce noise and readability of the structure factors, we apply some
data processing. We follow the same techniques used by Bergman \emph{et
al.}\cite{Bergman:2010wb}. Essentially along each constant $k$-vector, the data
is first smoothed with a Gaussian convolution of width ~0.95 THz 
and then normalised so that the maximum value is unity. This means that
the mode amplitudes can be compared only on constant $k$-vectors but not
between $k$-vectors. This is however, the more useful comparison, especially
when looking at the intermediate structure factors. Without normalisation,
the large change in temperature across the ISF would make the difference in
contrast between low and high temperature too large to reasonably display.
Further more we do no wish to compare the absolute value of the amplitude
at different times, but rather see which modes are more populated at each
given instance in time. An example of the Gaussian convolution is given in
figure~\ref{fig:convolution_example} showing that the data is smoothed but the
features remain intact.

\subsection*{S3 - Cluster Counting and Percolation Theory}

\begin{figure}[htbc!]
\centering
	\includegraphics[]{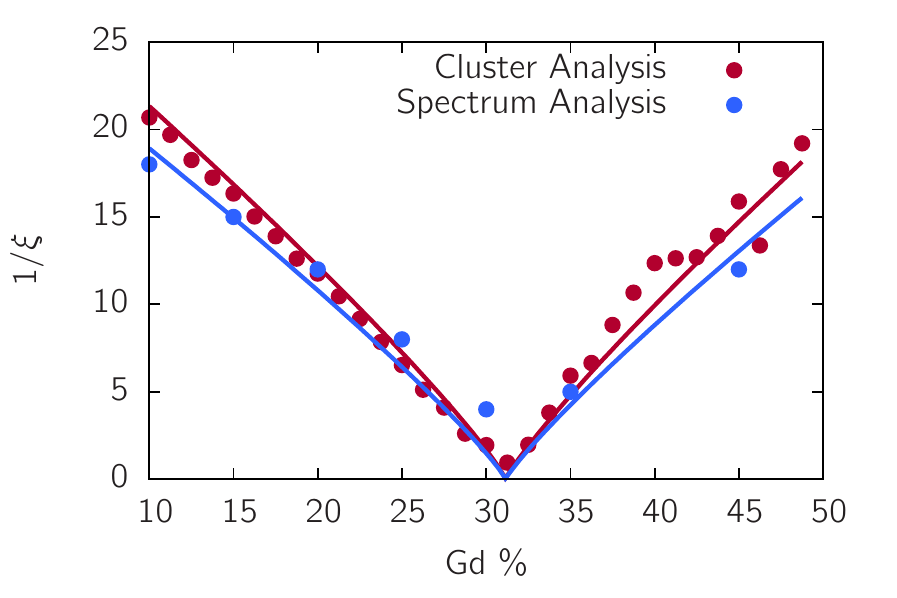}
    \caption{The red points correspond to data taken from cluster analysis using the Hoshen-Kopelman method on the lattice. The blue points are the maximum extent of the two spin wave mode as measured from the Langevin-Landau-Lifshitz-Gilbert dynamic structure factors (see Figure 2 in the main article). The lines correspond to the fit of Eq.~(\ref{eq:finite_size_scaling}) where only $A$ is a free parameter. For the cluster analysis $A=0.776187\pm0.01142$, analysis from the DSF gives $A=0.869468\pm0.04154$.}
  \label{fig:percolation}
\end{figure}

To identify clusters of Gd sites in the lattice we use the Hoshen-Kopelman
method\cite{Hoshen:1997ga}. This is an efficient algorithm for identifying
unique clusters on a lattice. We define a unique cluster as
any set of Gd sites which are linked together by a immediate adjacent site
(i.e. nearest neighbour exchange coupled). To calculate the typical correlation
length, we first remove the tails of the distribution \cite{Stauffer:1994uj},
that is any cluster of size $s<4$ and the percolating cluster. In practice we
discount the single largest cluster to avoid having to calculate if a cluster
is percolating. For the calculation of the correlation length, $\xi$ we use the
formula\cite{Stauffer:1994uj}
\begin{equation}
\xi^{2} = \frac{2\sum_{s}s^{2}\sum_{t=1}^{n_{s}}R_{st}^2}{\sum_{s}s^{2}n_{s}}	
\end{equation}
where $s$ is the cluster size, $R$ is the radius of gyration of a cluster and
$n_{s}$ is the number of clusters of size $s$. Universal critical exponents are strictly speaking defined only in the thermodynamic limit. Their determination from finite size lattice simulations requires performing the finite size scaling analysis\cite{Stauffer:1994uj}. However, considering very large lattices in our simulations reduces finite size effects and allows direct determination of critical exponents by fitting to
the lattices we generate using the form
\begin{equation}\label{eq:finite_size_scaling}
    \tilde{\xi} = A|p-p_{c}|^{-\nu}
\end{equation}
where $p_{c}$ is the site percolation threshold which is $0.3116004$ for a simple 
cubic lattice \cite{Grassberger:1992te} and the critical exponent
$\nu$ is $0.875$ (ref. \onlinecite{Lorenz:1998jm}) and $A$ is the only fitted parameter. The resulting fit is shown in
figure~\ref{fig:percolation} along with data extracted directly from the DSF as
the maximum length scale of the two-spin region. These show a good agreement as
discussed in the article.

\subsection*{S4 - Ferrimagnet Linear Spin Wave Theory}
To calculate the LSWT for arbitrary Gd compositions we use the spin analogy of the virtual crystal approximation to transform the disordered lattice 
Hamiltonian $\mathcal{H}$ describing the system,  to a translationally symmetric Hamiltonian $\mathcal{H}_{\mathrm{VCA}}$  with respect to spin variables $\mathbf{S}_i$.  This involves weighting the material parameters by the relative composition

\begin{align*}
    J_{0,\mathrm{11}} &= (1-x) z J_{\mathrm{11}}\\
    J_{0,\mathrm{22}} &= x z J_{\mathrm{22}}\\
    J_{0,\mathrm{12}} &= x z J_{\mathrm{12}}\\
    J_{0,\mathrm{21}} &= (1-x) z J_{\mathrm{21}}
\end{align*}
where subscripts $1$ and $2$ denote two different species, $x$ is the
concentration of species $2$, $z$ is the coordination of the lattice. 
The spin dynamics is described by the  Landau-Lifshitz equation of motion, 
$\mathrm{d} \mathbf{S}_i/\mathrm{d}t=\gamma[\mathbf{S}_i\times \mathbf{H}_{\mathrm{eff},i}]$, where $\mathbf{H}_{\mathrm{eff},i}=-\partial \mathcal{H}_{\mathrm{VCA}}/\partial \mathbf{S}_i$. 
The LL equation is  then transformed in terms of 
spin raising and lowering  operators $S^{\pm}_{i,\mathrm{FeCo}}=S_{i,\mathrm{FeCo}}^{x}\pm S_{i,\mathrm{FeCo}}^{y}$ which describes the spin fluctuations around equilibrium, in this case the $z-$ axis.  
The resulting system of two coupled equations is then 
Fourier transform to describe the spin fluctuations in the reciprocal space

\begin{equation}
\frac{\textrm{d}}{\textrm{d}t}\left(\begin{array}{c}
s_{\mathbf{k}1}^{+}\\
s_{\mathbf{k}2}^{+}
\end{array}\right)=-i\left(\begin{array}{cc}
\mathcal{A}_{\mathbf{k}11} & \mathcal{B}_{\mathbf{k}12}\\
\mathcal{B}_{\mathbf{k}21} & \mathcal{A}_{\mathbf{k}22}
\end{array}\right)\left(\begin{array}{c}
s_{\mathbf{k}1}^{+}\\
s_{\mathbf{k}2}^{+}
\end{array}\right)\label{eq:GeneralMatrixEquation}
\end{equation}
where
\begin{align*}
    \mathcal{A}_{\mathbf{k}11}&=\frac{\gamma}{\mu_{1}}\left(J_{0,11}-J_{\mathbf{k}11}\right)\langle s_{1} \rangle+\frac{\gamma}{\mu_{1}}J_{0,12}\langle s_{2} \rangle \nonumber \\
    \mathcal{B}_{\mathbf{k}12}&=\frac{\gamma}{\mu_{1}}  J_{\mathbf{k}12}\langle s_{2} \rangle\nonumber \\
    \mathcal{A}_{\mathbf{k}22}&=\frac{\gamma}{\mu_{1}}\left(J_{0,22}-J_{\mathbf{k}22}\right)\langle s_{2} \rangle+\frac{\gamma}{\mu_{1}}J_{0,21}\langle s_{1} \rangle\nonumber \\
    \mathcal{B}_{\mathbf{k}21}&=\frac{\gamma}{\mu_{2}}  J_{\mathbf{k}21}\langle s_{1}. \rangle
\end{align*}
We  include the temperature in a mean field approximation by the self consistent
calculation of the values of $\langle s_{1} \rangle$ and $\langle s_{2}
\rangle$ \cite{Ostler:2011jf}.
Matrix equation \eqref{eq:GeneralMatrixEquation} is   
  diagonalised using the  Bogoliubov-like transformation   $\mathbf{s}_{\mathbf{k},\mathrm{FeCo}}=u^+_{\mathbf{k}}\alpha_{\mathbf{k}}+u^{-}_{\mathbf{k}}\beta^{\dag}_{\mathbf{k}}$.
$\mathbf{s}_{\mathbf{k},\mathrm{Gd}}=u^-_{\mathbf{k}}\alpha^{\dag}_{\mathbf{k}}+u^{+}_{\mathbf{k}}\beta_{\mathbf{k}}$, where $\alpha_{\mathbf{k}}$
and $\beta_{\mathbf{k}}$  
are the eigenstates (magnons) of the system with  frequency

\begin{align}
\omega_{\alpha}(\mathbf{k}) & =\frac{1}{2}\left[\sqrt{\left(\mathcal{A}_{\mathbf{k}11}+\mathcal{A}_{\mathbf{k}22}\right)^{2}-4\mathcal{B}_{\mathbf{k}12}\mathcal{B}_{\mathbf{k}21}}-\left(\mathcal{A}_{\mathbf{k}22}-\mathcal{A}_{\mathbf{k}11}\right)\right]\\
\nonumber \\
\omega_{\beta}(\mathbf{k}) & =
\frac{1}{2}\left[\sqrt{\left(\mathcal{A}_{\mathbf{k}11}+\mathcal{A}_{\mathbf{k}22}\right)^{2}-4\mathcal{B}_{\mathbf{k}12}\mathcal{B}_{\mathbf{k}21}}-\left(\mathcal{A}_{\mathbf{k}11}-\mathcal{A}_{\mathbf{k}22}\right)\right]
\end{align}
where the coefficients of the transformation $u_{\mathbf{k}}$ and $v_{\mathbf{k}}$ read

\begin{align}
u_{\mathbf{k}} & =\sqrt{\frac{1}{2}\left(\frac{\mathcal{A}_{\mathbf{k}11}-\mathcal{A}_{\mathbf{k}22}}{\sqrt{\left(\mathcal{A}_{\mathbf{k}11}+\mathcal{A}_{\mathbf{k}22}\right)^{2}-4\mathcal{B}_{\mathbf{k}12}\mathcal{B}_{\mathbf{k}21}}}+1\right)}\\
\nonumber \\
v_{\mathbf{k}}& =\sqrt{\frac{1}{2}\left(\frac{\mathcal{A}_{\mathbf{k}11}-\mathcal{A}_{\mathbf{k}22}}{\sqrt{\left(\mathcal{A}_{\mathbf{k}11}+\mathcal{A}_{\mathbf{k}22}\right)^{2}-4\mathcal{B}_{\mathbf{k}12}\mathcal{B}_{\mathbf{k}21}}}-1\right)}
\end{align}

\newpage 
\section*{Additional Information}
\subsection*{Acknowledgements}
This work was supported by the EU Seventh Framework Programme under grant
agreement No. 281043, FEMTOSPIN. and the Spanish project from Ministry of
Science and Innovation under the grant FIS2010-20979-C02-02.  U.A. is funded by
Basque Country Government under "Programa Posdoctoral de perfeccionamiento de
doctores del DEUI del Gobierno Vasco".  O.H. gratefully acknowledges support
from a Marie Curie Intra European Fellowship within the 7th European Community
Framework Programme under Grant Agreement No. PIEF-GA-2010- 273014
(MENCOFINAS).

\subsection*{Author Contributions}

J.B. and T.O. performed atomistic spin dynamics simulations; J.B., U.A. and O.F. performed the LSWT calculations and interpretation and J.B., O.H. and R.C. carried out the HK cluster analysis and percolation theory. J.B. and U.A. wrote the core of the manuscript, all authors contributed to parts of it.

\subsection*{Competing Financial Interests}
The authors declare no competing financial interests.
\newpage

\section*{Figure Legends}

%=======================================================
%=========================================================

%=========================================================
%=======================================================

%=======================================================

%=======================================================
%=======================================================

%=======================================================

\end{document}